# Cloaking and imaging at the same time

Qiannan Wu *, Yadong Xu *, and Huanyang Chen *, †

*School of Physical Science and Technology, Soochow University, Suzhou, Jiangsu 215006, China*

In this letter, we propose a conceptual device to perform subwavelength imaging with positive refraction. The key to this proposal is that a drain is no longer a must for some cases. What's more, this device is an isotropic omnidirectional cloak with a perfect electric conductor hiding region and shows versatile illusion optical effects. Numerical simulations are performed to verify the functionalities.

* These authors contribute equally.
† chy@suda.edu.cn

Perfect lens [1,2] and invisibility cloak [3,4], the two attractive concepts, can be both derived from transformation optics [3,4,5]. "The cloak can hide things, but the perfect lens can see anything." [6] There seems to be a big gap between them [7]. In literature, perfect lens (with negative refraction) [1] was found to be able to hide things [8], yet it seems unable to perform imaging at the same time [9]. In this letter, we will show that one kind of cloak - the conformal cloak [3] can be used not only to hide things but also to perform imaging. The mirrored Maxwell's fish eye lens was found before [2] to make perfect lens if a perfect drain is applied at the imaging point [10] (although it is still a controversial claim [11,12,13,14,15] and is regards to be similar to the super-resolution image schemes from time reversal [16,17]). This time, again with the aid of the mirrored Maxwell's fish eye lens, we design an omnidirectional cloak with a positive refractive index profile and a perfect electric conductor hiding region. The branch cut of the conformal cloak will be proved later to serve as a perfect drain, consequently enabling the device to perform subwavelength imaging without any external drain.

Let's start from the Zhukowski mapping that inspires the conformal cloak [3],

$$w = z + \frac{a^2}{z}. \tag{1}$$

The mapping sets up a correspondence between two spaces: the $z$-space (physical space) and the $w$-space (virtual space). The $w$-space has two Riemann sheets, one (the lower sheet) mapped to the interior region ($|z|<a$) of the $z$-space, and the other (the upper sheet) mapped to the exterior region ($|z|>a$). The branch cut (a segment from $w_2 = -2a$ to $w_1 = 2a$) is mapped to the boundary $|z|=a$ (a circle). The conformal mapping establishes the relationship between the refractive index profiles in the two spaces:

$$n_z = n_w |\frac{dw}{dz}|. \qquad (2)$$

Now let's suppose $n_w = 1$ in both sheets. When light rays in the upper sheet don't cross the branch cut, they will stay in the same sheet in the virtual space. In the physical space, they will stay in the exterior region. When they cross the branch cut, they will go into the lower sheet. Physically, they will cross the circular boundary and go into the interior region. Then how to build up an invisible cloak? A specific refractive index profile should be applied in the lower sheet so that whenever light rays cross the branch cut, they will propagate into the lower sheet but in closed orbits and cross the branch cut again to go back to the upper sheet. As light rays cannot reach a region in the lower sheet, physically it is mapped to the invisible region. Fox example, if we set,

$$n_w = \begin{cases} \dfrac{2}{1+\dfrac{|w-2a|^2}{4a}}, & |w-2a| \leq 4a, \\ \infty, & |w-2a| > 4a. \end{cases} \qquad (3)$$

This mirrored Maxwell's fish eye lens will force all the light rays going into the lower sheet to propagate in closed orbits and go back to the upper sheet (see in Fig. 1(a)). As the rays cannot reach the exterior region which is outside a perfect electric conductor (PEC) boundary ($|w-2a|=4a$) in the lower sheet, this region is mapped to a PEC region that can hide things inside (see in Fig. 1(b)). Such a cloaking effect is valid in

geometrical optics. For some quantized wavelengths, the effect is nearly perfect in wave optics [18]. For this profile, if

$$\frac{8\pi a}{\lambda} = \sqrt{l(l+1)}, \tag{4}$$

the cloaking effect still happens. To show this, we perform a numerical simulation in Fig. 1(b) for $l = 24$. The scattering pattern confirms good cloaking effect. Note that a cut-off radius is introduced ($r_c = 5a$) for the exterior region during simulation (outside the cut-off radius, the material is replaced with air as an approximation) [18]. This type of conformal cloak has an advantage compared to that in Ref. [18]. That is, the cloaking region is a PEC, hence easy to be utilized in hiding objects. In addition, compared to the original push-forward mapping cloak [4,19] and non-Euclidean cloak [20,21], the cloak here has isotropic material parameters and only requires a refractive index profile to perform good cloaking effect. However, the present conformal cloak is only limited to two dimensional operation, while the push-forward mapping cloak and non-Euclidean cloak have an advantage of working in three dimension. Another drawback is that the refractive indexes here have a maximum value of 33 approximately, which is regarded too large even for implementation in microwave frequencies.

In the following, we will consider another kind of profile in the lower sheet:

$$n_w = \begin{cases} \dfrac{2}{1+\dfrac{|w-2a|^2}{2a}}, & |w-2a| \le 2a, \\ \dfrac{2}{1+\dfrac{|w+2a|^2}{2a}}, & |w+2a| \le 2a, \\ \infty, & others. \end{cases} \tag{5}$$

It is two kissing mirrored Maxwell's fish eye lenses. Such a profile can guide light rays to propagate in closed orbits in the lower sheet (see in Fig. 1 (c)), leading to the same cloaking effect as that from Eq. (3). Moreover, the maximum refractive index has a value of about 13, which is almost one third of the above-mentioned. If we change the above mirrored Maxwell's fish eye lenses into Eaton lenses [18], the maximum refractive index will be reduced to 5.862 (not shown here). Therefore, we believe that by searching and changing the geometry and refractive index profile in the lower sheet, the maximum value could be lowered to a more desirable range (e.g., less than 4) for future implementations using metamaterials. The cloaking effect is also valid for both geometry optics and wave optics. The quantized wavelengths for Eq. (5) should be changed into:

$$\frac{4\pi a}{\lambda} = \sqrt{l(l+1)}, \tag{6}$$

Figure 1 (d) is a numerical simulation for $l=12$, demonstrating the cloaking effect very well. The profile described in Eq. (5) shows a more suitable refractive index range, laying some ground for further implementations.

In the next section, we will prove that the above devices can even perform sub-wavelength imaging functionality. We will only focus on the geometry in Fig. 1(a). Figure 1 (c) will also show similar functionality. As the principle is the same, related results would not be shown here. Figure 2 (a) is a schematic plot to demonstrate our idea. Suppose we have an active line current source at the segment from $w=2a$ to $w=6a$ in the lower sheet. The mirrored Maxwell's fish eye lens can perform perfect imaging, if there is a drain at the segment from $w=-2a$ to $w=2a$. We then have a speculation. As all the light rays that cross the branch cut from the lower sheet will go into the upper one, the branch cut may naturally serve as a perfect drain. If so, a perfect image can be observed in the upper sheet. If there are more than two sources at the segment from $w=2a$ to $w=6a$, the branch cut can also serve as drains and project the images to the upper sheet.

Let us see the case for one active source firstly. If there is a line current source at $w=4a$, the image position should be at $w=0$. In the physical space, a line current source is applied at $z=(2-\sqrt{3})a$. It forms two images at $z=\pm i$. For the far field pattern, it looks as if there is only one source (image) at $z=0$, due to the transformation medium outside $|z|=a$ ($n_z=|1-\frac{a^2}{z^2}|$). Only at the quantized wavelengths described in Eq. (4) can the conformal cloak inherit the perfect imaging functionality from geometry optics to wave optics. Therefore we choose the same wavelength in Fig. 1(b) for instance ($l=24$). Figure 2(b) is the simulation result well proving that the branch cut can serve as a drain. Figure 2(c) is the enlarged plot of 2(b). The power outflow is also plotted in Fig. 2(d) as a function of $\theta$ at the circular boundary. The resolution can be estimated to be less than one third of the wavelength, demonstrating that the image is a subwavelength one. Numerically, we also find that, for an even $l$, the device looks like a source at the origin but with an opposite phase, while for an odd $l$, the image has the same phase as that of the source.

Then we move to the case for two or more sources. We will see that such a device can help identify sources in a line and perform a very sub-wavelength imaging functionality. For example, we have two line current sources (in-phase here, could be arbitrary), one at $w = 4a$, and the other at $w = 5a$ (physically, their distance is less than $0.06a$). The image positions should be at $w = 0$ and $w = -a$, respectively. Figure 3(a) is the electric field distribution demonstrating that the far field pattern looks like two interfering sources in vacuum (but both with opposite phases to those of the original sources), as shown in fig. 3(b). From Fig. 3(a), we can see there are four images of the two sources at the circular boundary, clearly demonstrating the subwavelength imaging functionality. We enlarge the plot of Fig. 3(a) and show it in Fig. 3(c) for an evident view of the imaging functionality. Figure 3(d) is the power outflow as a function of $\theta$ at the circular boundary, where two subwavelength images with a resolution of less than one third of the wavelength are clearly shown.

Apart from the above interesting functionalities, the device is even more versatile. For example, if we put two in-phase line current sources at $z = \pm i$, the far field pattern will look like only one source at the origin but with twice intensity, as shown in Fig.

4(a). Such effect resembles one of the properties of the conformal lenses [22]. If we put two out-of-phase line current sources at $z = \pm i$, the intensity of the outcoming waves will diminish significantly compared to that of the active sources, as shown in Fig. 4(b). In other words, the device can also be utilized to cloak active sources, which often negative index materials were used [23].

In conclusion, we found that mirrored Maxwell's fish eye lens can be used to design a conformal cloak with a PEC cloaking region for both geometry optics and wave optics. Two kissing mirrored Maxwell's fish eye lenses can even help design a cloak with a more desirable refractive index profile. The device can also perform sub-wavelength imaging, inheriting an advantage that only isotropic positive index materials are required. At the same time, the drains need not to be applied inside the device as the branch cut itself serves as drains. Finally, the device can also perform some illusion optical effects, such as making two sources like one and cloaking an active source. All in all, the device is important at least in two aspects. One is that it is an isotropic cloak with a PEC hiding region for all directions. The other is that drains are not necessary for some cases of sub-wavelength imaging, for example, when figuring out sources in a line ("no drain, but gain" [15]).

This work was supported by the National Natural Science Foundation of China (grant no. 11004147), the Natural Science Foundation of Jiangsu Province (grant no. BK2010211) and the Priority Academic Program Development (PAPD) of Jiangsu Higher Education Institutions.

Fig. 1 (Color online) Cloaking effect in geometry optics and wave optics. (a) A mirrored Maxwell's fish eye lens applied in the lower sheet can make light rays propagating from the upper sheet (the blue/dashed lines) go in closed orbits (black eye-shaped curves) and turn back to the upper sheet again. (b) The related cloaking effect of (a) for waves, performed numerically for $l = 24$ in Eq. (4). (c) Two kissing mirrored Maxwell's fish eye lenses can function similarly as that in (a) and construct a conformal cloak with a more suitable refractive index profile. (d) The related cloaking effect of (c) for waves, performed numerically for $l = 12$ in Eq. (6).

Fig. 2 (Color online) Imaging effect in geometry optics and wave optics. (a) The mirror Maxwell's fish eye lens can make perfect lens if a drain is applied at the image point. As all the light rays crossing the branch cut will propagate from one Riemann sheet to another, we speculate that the branch cut itself serves as a drain if the source is suitably located. For example, the source in the lower sheet (black empty circle) can form an image at the branch cut (red point), and eventually the image will be transferred to the upper sheet. (b) The electric field distribution when an active source is applied in the conformal cloak. The device can project the source from the inner region to the exterior region. (c) The enlarged plot of (b), where the source is denoted by an 'x' sign. (d) The power outflow at the circular boundary from $\theta = 0$ to $\theta = \pi$.

Fig. 3 (Color online) The subwavelength imaging functionality of the conformal cloak.

(a) The device can be used to identify two active sources located in a very sub-wavelength distance, and make them look like two sources in vacuum interfering with each other. The electric field distribution is plot in (b). (c) The enlarged plot of (a), where the sources are denoted by two 'x' signs. (d) The power outflow at the circular boundary from $\theta = 0$ to $\theta = \pi$.

Fig. 4 (Color online) The illusion optical effects of the conformal cloak. (a) The device can make two active sources into one. (b) The device can cloak one active source.

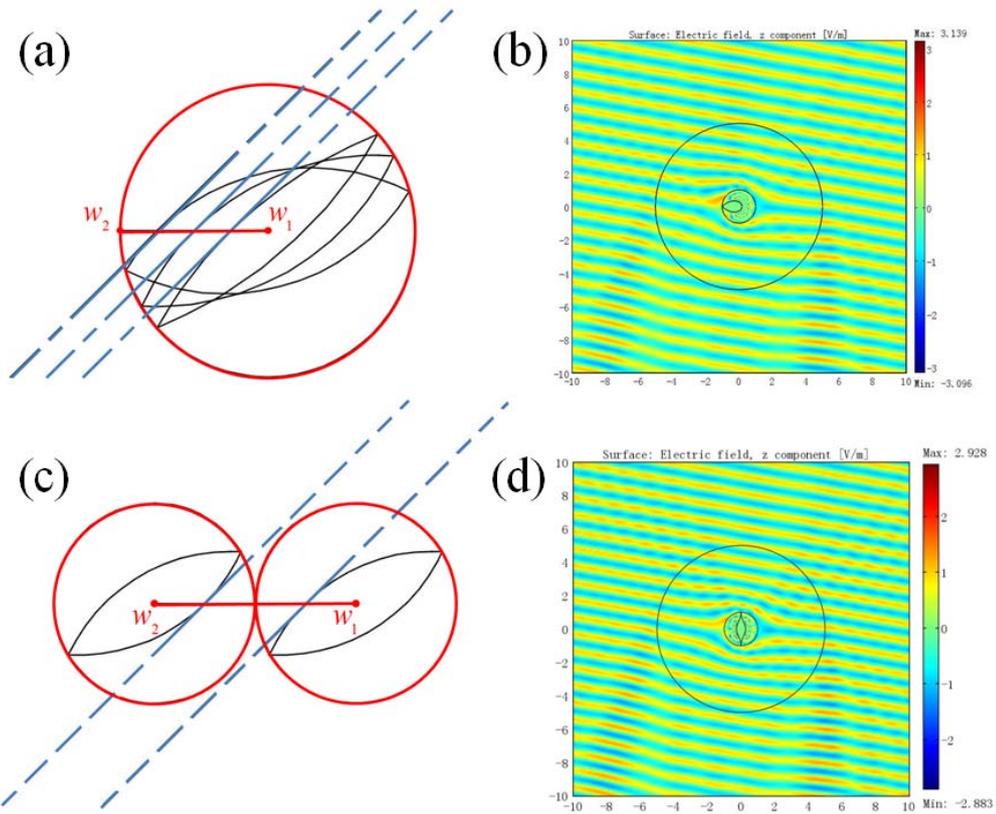

Fig. 1

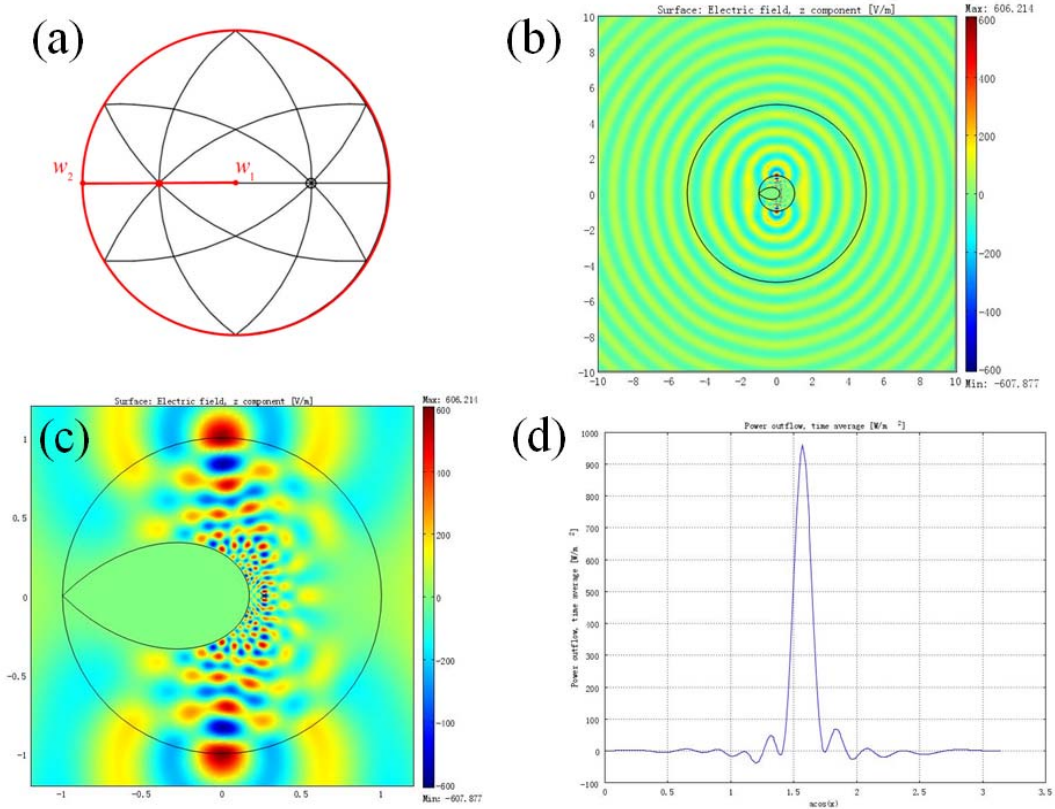

Fig. 2

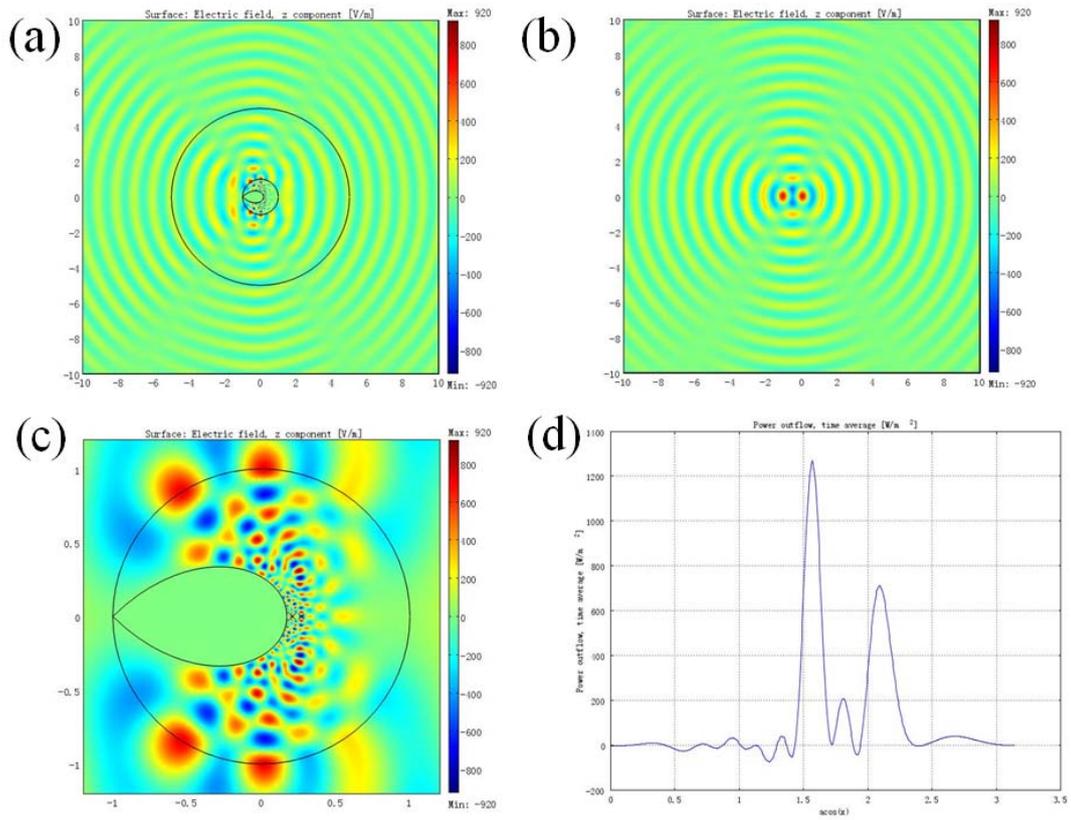

Fig. 3

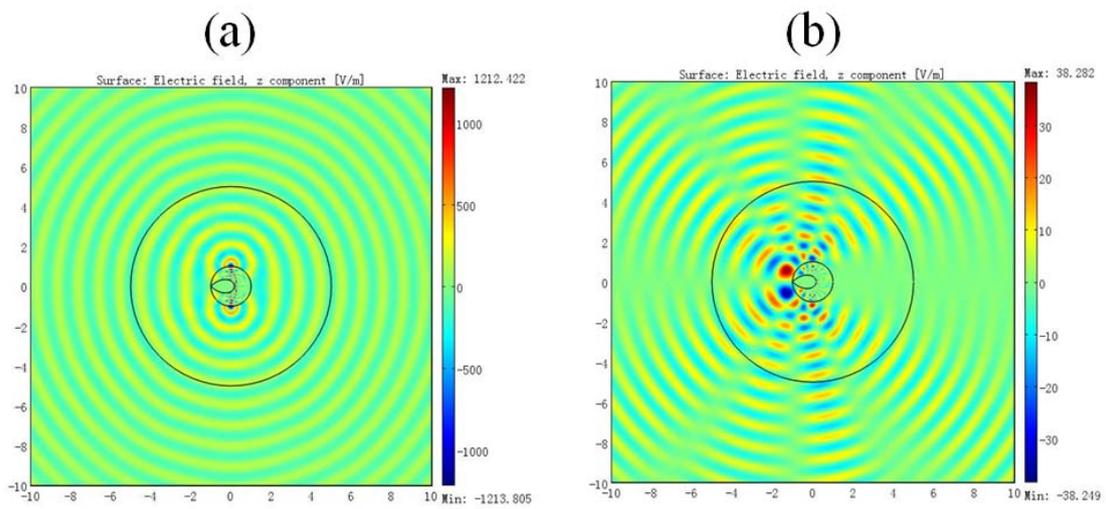

Fig. 4